\documentclass[a4paper, amsfonts, amssymb, amsmath, preprint, showkeys, nofootinbib, twoside,superscriptaddress]{revtex4-2}
\usepackage[english]{babel}
\usepackage[utf8]{inputenc}
\usepackage{gensymb}
\usepackage{lineno,soul,multirow}
\usepackage[utf8]{inputenc}
\usepackage{graphicx,color,float} 
\usepackage[pdftex, pdftitle={Article}, pdfauthor={Author}]{hyperref} 
\begin{document}
\title{Phase stability and defect studies of Mg-based Laves phases using defect phase diagrams }

\author{A. Tehranchi}
    \affiliation{Max-Planck-Institut f\"ur Eisenforschung GmbH, D-40237 D\"usseldorf, Germany  }

\author{M. Lipinska-Chwalek}
    \affiliation{Ernst Ruska-Centre for Microscopy and Spectroscopy with Electrons (ER-C), 52428 Forschungszentrum J\"ulich, Germany  }

\author{J. Mayer}
    \affiliation{Ernst Ruska-Centre for Microscopy and Spectroscopy with Electrons (ER-C), 52428 Forschungszentrum J\"ulich, Germany  }

\author{ J. Neugebauer}
    \affiliation{Max-Planck-Institut f\"ur Eisenforschung GmbH, D-40237 D\"usseldorf, Germany  }

\author{T. Hickel}
    \affiliation{Max-Planck-Institut f\"ur Eisenforschung GmbH, D-40237 D\"usseldorf, Germany  }
     \affiliation{BAM Federal Institute for Materials Research and Testing, D-12489 Berlin, Germany}

\date{\today} 
\begin{abstract}
Laves phases often form as secondary phases in metallic alloys and have a significant effect on their structural properties.  Thus, phase stability studies for these chemically and structurally complex phases in addition to mechanical behavior studies are of great interest.  In this work, we use the concept of metastable bulk phase and defect phase diagrams to augment the understanding of the bulk phase and defect phase stability in Laves phases in  Mg-based alloys. In this way, we resolve the discrepancy between bulk phase diagrams and experimental observations regarding the formation of Mg-rich C14 and Al-rich C15 Laves phases in MgAlCa alloys at moderate temperatures.  Moreover, the effect of the thermodynamic state of alloys on the competition between solute-rich hcp-like planar defects and stoichiometric basal stacking faults is clarified, which determines the brittleness of these alloys.  \end{abstract}


\maketitle

 \section{Introduction}\label{Sec:intro}
 Laves phases are intermetallic compounds that form in various metallic alloys and are often crucial for their structural performance.  They exist in two hexagonal crystal structures, C14 and C36, in addition to the cubic structure C15.  The prototypes of these phases are Zn$_2$Mg, Ni$_2$Mg, and Cu$_2$Mg, respectively~\cite{bragg1931strukturbericht}. Since the presence of these phases strongly depends on the thermo-mechanical treatment of the alloys, it is necessary to study the impact of the chemical composition and defects on the thermodynamics of their formation. 
 
 The material system discussed in this paper is formed by Mg-based alloys. They are highly attractive for applications due to their low cost and weight~\cite{zubair2021co,guenole2021exploring}.  The main alloying elements in this material system are Al, Zn, and Mn~\cite{wu2018mechanistic,guenole2021exploring}.  Moreover, the addition of Ca  usually enhances their high-temperature mechanical properties \cite{sandlobes2017rare} and leads to the formation of CaMg$_x$Al$_{2-x}$ Laves phases. 
The phase diagrams suggest that at low and moderate temperatures the stable Laves phases are Mg-rich C14 and Al-rich C15, whereas C36 is only stable for intermediate Mg concentrations, i.e., $0.5\le x\le1.5$, at high temperatures~\cite{kevorkov2010400,zhong2006al2}.  Our experiments show, however, that the Al-rich C15 phase doesn't form in some casts of Mg-based alloys with a considerable amount of Al despite the prediction of the bulk phase diagrams at lower temperatures.  This observation is due to the high nucleation barrier for the cubic C15 phase in hexagonal Mg, and C36 phases.  The information about this barrier is absent in the bulk phase diagrams.  Moreover, interestingly in one of our experimental samples  Mg-rich precipitates form within the C36 phase.  The presence of these precipitates results in formation of entirely different planar defects in these samples rather than usual stacking faults which form in the other samples after deformation.   This change in the phases of the planar defects is crucial since these defect dictate the response of the material to external loads during its lifetime.

In this work we employ the concept of the metastable phase diagram (MPD) and  metastable defect phase diagram (MDPD), to resolve the limitations of the bulk phase diagrams in the prediction of the formation of bulk phases as well as rationalization of the formation of the different defect phases in our experimental alloys.  For a thorough introduction of the concepts behind these diagrams see Ref.~\cite{tehranchi2023metastable}.  
The remainder of this paper is organized as follows.  Section~\ref{sec:exp} contains the experimental procedures and observations.  Stability of the bulk phases as well as the defect phases are given in Section~\ref{sec:stable} and Section~\ref{Sec:conc} contains the conclusions of this work. 

\section{Experimental observations of planar defects in the Mg-Ca-Al system}\label{sec:exp}
In this work, we study two different composite materials (S1 and S2), consisting of the $\alpha$-Mg matrix and the C36 Laves phase. Both alloys were cast and deformed in a very similar way and possess a nominal composition close to Mg-5Al-3Ca (wt.\%).  
The S1 material is the Mg-5Al-3Ca alloy described in Ref.~\cite{zubair2019role}, while S2 is the sample that was used in Ref.~\cite{guenole2021exploring}. The chemical composition of the alloys and of the phases observed in both samples are summarized in Table SI of the Supplementary document. 

Fig.~\ref{Fig:exp} shows the overview images of the C36 Laves phase skeleton of the samples S1 and S2 (a and d, respectively) with corresponding high-resolution details of the planar defects observed in the particular sample. 
The planar defects are only formed after application of 5\% uniaxial strain at 170$^{\degree}$C.  Details of the structure and deformation behavior and detailed atomic description of the observed structures is a subject of a separate paper, but particular difference between the two materials can be observed directly in HAADF images (Figs.~\ref{Fig:exp}a and d). 

The presence of homogeneously distributed nanoprecipitates (indicated with red arrows), coherent with the surrounding C36 structure, together with a high amount of evenly distributed planar defects (indicated with blue arrows), passing the hexagonal C36 Laves phase (diagonally in HAADF image) along its basal planes, is characteristic for the S1 alloy. 
Corresponding EDX distribution maps for Mg, Al, and Ca, reveal a Mg enrichment of the precipitates and planar defects observed in alloy S1 (Fig.~\ref{Fig:exp}g). 

While the nominal composition of S2 is only slightly different from that of S1, neither the presence of the Mg-rich precipitates nor a Mg-enrichment of the deformation defects was found in the S2 material (see EDX maps in Fig.~\ref{Fig:exp} f). It is also worth to note that the deformation of C36 in the S2 alloy is significantly more localized and results in evident steps at the interface between Mg matrix and the deformed C36 Laves Phase (as indicated by blue arrow in Fig.~\ref{Fig:exp} d). 
Moreover, the atomic structure of the deformation planar defects differs significantly between the two investigated alloys. The Mg-rich planar defects in S1 material do not introduce any stacking fault (SF) in the surrounding C36. This can be recognized by the perfect match between the crystal structure on both sides of the defect (Fig.~\ref{Fig:exp}b), indicated with the help of the simplified tiling. Orange, green, yellow and blue tiles represent the main building blocks of the hexagonal C36 Laves phase, as observed along the $[11\bar{2}0]$ crystallographic direction (the distance measured along $[0001]$ direction from the center of one orange tiling, via green, yellow and blue tiling, to the center of the next orange tiling represents the height of the single hexagonal C36 unit cell). 

Contrary to the S1 material, planar defects observed in the S2 material involve the presence of various SFs and related dislocations at the end of the planar defect. 
For example, an exemplary defect shown in Fig.~\ref{Fig:exp}e involves displacements related to the slip by 1/3$[01\bar{1}0]$ combined with an extrinsic SF and related structure displacement by 1/4[0001].
The Fourier transform shown in Fig.~\ref{Fig:exp}c was obtained from high resolution images of the S1 sample within the areas indicated in Fig.~\ref{Fig:exp}a with small squares 1, 2 and 3 (1-within C36 region 2 -inside the Mg-rich precipitate and 3- from the region containing both C36 and precipitate structure) and reveals the crystal structure (hcp) of the Mg-rich precipitate and the orientation relationship between the precipitates and the surrounding C36 phase ($[11\bar{2}0]$C36$\parallel$$[01\bar{1}0]$hcp prec. and $[0001]$C36$\parallel$ $[0001]$ hcp precipitate).

\begin{figure}[ht]
         \centering
         \includegraphics[width=\textwidth]{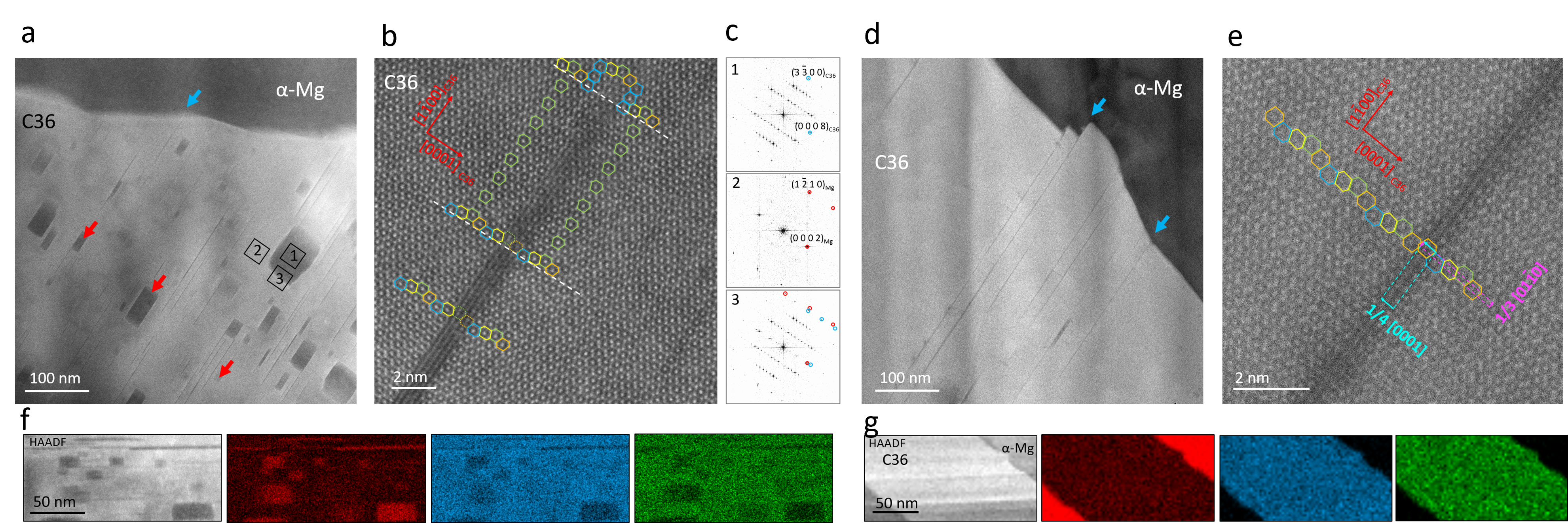}
         \caption{STEM images of the structure observed in the S1 (a-c and f) and S2 sample (d, e and g). The overview HAADF images reveal the structures characteristic for S1(a) and S2 (d) material, respectively. The deformation planar defects are indicated with blue arrows in (a) and (d), while Mg-rich precipitates are indicated with red arrows in (a). High resolution HAADF images reveal the atomic structure of the C36 skeleton and the  deformation defects forming in the S1 and S2 alloy (b and e, respectively). Corresponding to the HADDF images, EDX distribution maps  of Mg (red), Al (blue) and Ca (green) are presented for the structures S1 (f) and S2 (g).}
         \label{Fig:exp}
     \end{figure}
     
\section{Stability of bulk and defect phases in Mg-based ternary alloys}\label{sec:stable}

\subsection{Formation energies of Laves phases in the ternary Mg-Ca-Al system}

\begin{figure}[ht]
 \includegraphics[width=\textwidth]{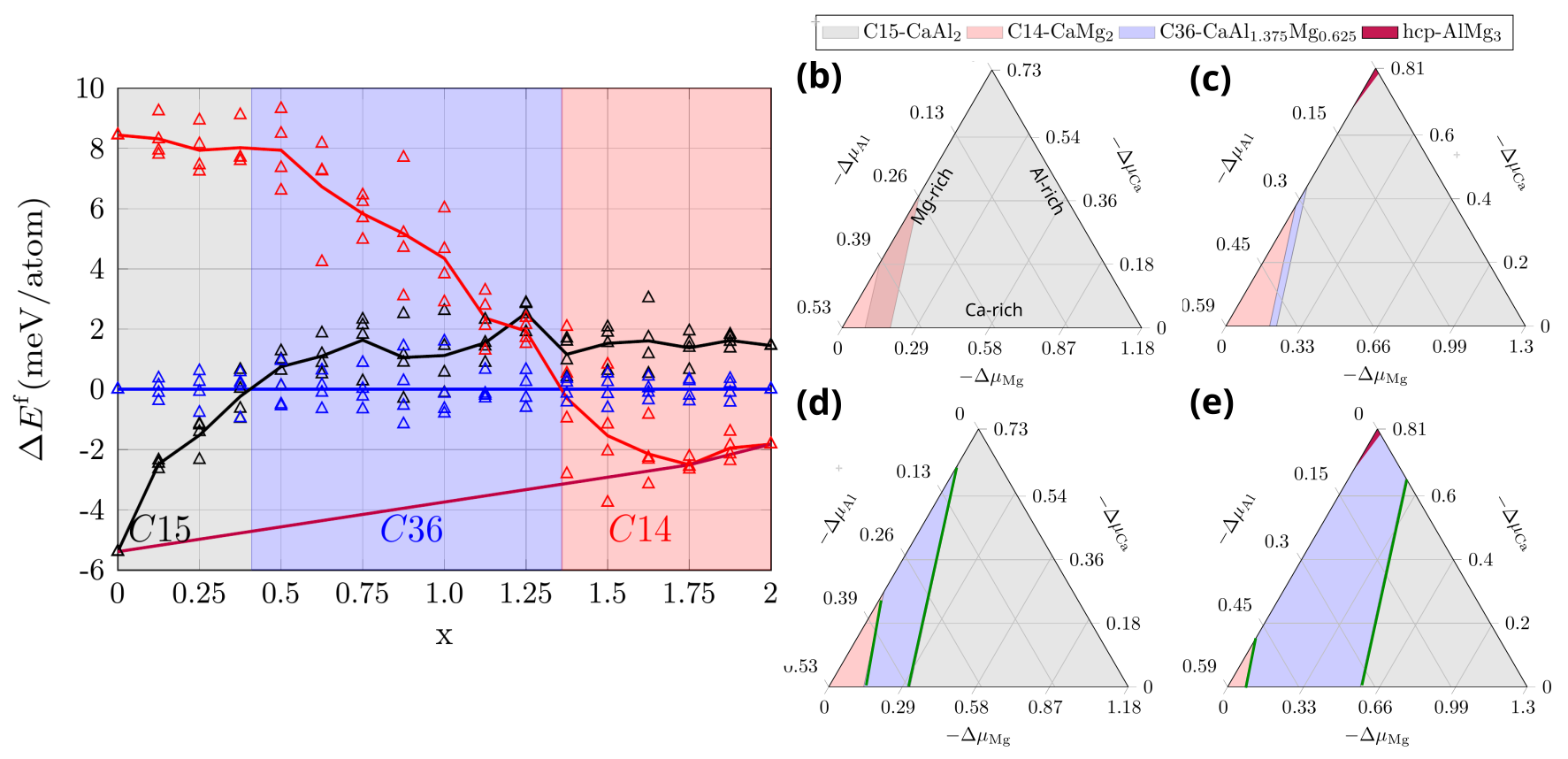}
    \caption{(a) Variation of the relative formation energies $\Delta E_{\rm f}(x)=E_{\rm f}^{\rm L}(x) - \bar E_{\rm f}^{\rm C36}(x)$ of different Laves phases with composition CaMg$_x$Al$_{2-x}$ versus $x$. Each symbol indicates the result of an independent DFT calculation. The color coding of the background illustrates the composition domains at which a certain Laves phase is most stable.    The convex hull of the average formation energies is given by the purple line. (b) and (c) show ternary bulk phase diagrams of Laves phases with $x$=0.625 at $T = 0$ and $T = 773$ K, respectively. (d) and (e) are ternary metastable phase diagrams of Laves phases with $x$=0.625 at $T = 0$, and $T = 773$ K, respectively.
    }
        \label{fig:DFTCross}
\end{figure}

In this section, the focus is on intermetallic phases in the ternary Mg-Ca-Al system, namely the  C14, C15, and C36 Laves phases for various compositions of $\rm{CaMg}_{x}\rm{Al}_{2-x}~(0\le \textit{x}\le 2)$. 
We performed DFT calculations to determine their geometry and formation energy. 
To statistically capture the chemical disorder, we consider four random atomic configurations for each composition and Laves phase L.  Fig.~\ref{fig:DFTCross}a shows these formation energies
\begin{align}
    E_{\rm f}^{\rm L}=\frac{1}{3}\left(E({\rm Ca}{\rm Mg}_{x}{\rm Al}_{2-x})-\mu^0_{\rm Ca}-x\mu^0_{\rm Mg}-(2-x)\mu^0_{\rm Al}\right)
\end{align}
with respect to the average formation energy of C36 phase at $x$, $\bar{E}_{\rm f}^{\rm C36}(x)$, where $E({\rm Ca}{\rm Mg}_{x}{\rm Al}_{2-x})$ is the energy per formula unit of the bulk Laves phase of interest, and the $\mu^0_{\rm X}$ are the chemical potentials of fcc-Ca, hcp-Mg, and fcc-Al, respectively.

The color coding of the background illustrates the composition domains at which a certain Laves phase is stable with respect  to the other Laves phases at \emph{fixed composition}. The convex hull of the mean values, which is also given in the plot, indicates that the C36 phase should decompose at $T=0$K into C14 and C15 for all compositions.  

\subsection{Ternary metastable phase diagrams}
In order to understand the experimentally observed planar defects in the the C36 Laves phase, we construct the ternary phase diagrams of the Laves phases 
${\rm CaMg}_x{\rm Al}_{2-x}$ using the relation
\begin{align}
    \Delta\mu_{\rm Ca}+x\Delta\mu_{\rm Mg}+(2-x)\Delta\mu_{\rm Al}=3E_{\rm f}({\rm CaMg}_x{\rm Al}_{2-x}), \label{Eq:ternary}
\end{align}
where $\Delta\mu_{\rm X}=\mu_{\rm X}-\mu^0_{\rm X}$. For example, Fig.~\ref{fig:DFTCross}b illustrates the ternary diagram constructed for C36-${\rm CaMg}_{0.625}{\rm Al}_{1.375}$ at $T = 0$ K.  This choice is close to the composition of the C36 phase in the experimental sample S1 and preserves its Mg/Ca ratio. Every point in the triangle corresponds to a set of chemical potentials that fulfills the condition in Eq.~\eqref{Eq:ternary} for $x = 0.625$. 
Each edge of the triangle corresponds to the condition $\Delta\mu_{\rm X}=0$, which means the formation of the bulk phase of atom X.  In the case of the above-described experiment the C36 phase is embedded in the Mg hcp matrix, thus the experimental condition implies $\Delta\mu_{\rm Mg}\approx 0$ and we are essentially back to a one-dimensional problem.   However, since the C36 phase is metastable at $T = 0$ K, we are now more careful with applying constraints of conventional thermodynamics. 

The stability domain for the C36 phase can be found using the formation energies of the end members for the decomposition (cf. Fig.~\ref{fig:DFTCross}a), i.e.,

\begin{align}
  \notag \Delta\mu_{\rm Ca}+2\Delta\mu_{\rm Mg}&\le3E_{\rm f}({\rm CaMg}_2)\\
    \Delta\mu_{\rm Ca}+2\Delta\mu_{\rm Al}&\le3E_{\rm f}({\rm CaAl}_2)\label{Eq:admiss}
\end{align}
Here, we have not considered the solubility of Al in C14 (cf.~Fig.~\ref{fig:DFTCross}), after having convinced ourselves that these corrections result into small quantitative, but no qualitative changes. At $T=0$ K this yields a coexistence of the C14 and C15 phase, but no stability of C36.

To include the effect of temperature one should consider the formation \emph{free} energy of the phase of interest at the right-hand side of Eq.~\eqref{Eq:ternary}. In this work, we consider the configurational contribution,
\begin{align}
    F_{\rm f}(T)=E_{\rm f}+\frac{2}{3}k_{\rm B}T\left(\frac{x}{2}\ln \frac{x}{2}+\frac{2-x}{2}\ln \frac{2-x}{2}\right),
\end{align}
where $k_{\rm B}$ is the Boltzmann constant. Using these corrections we constructed the ternary phase diagrams for a representative high temperature ($T=$ 773 K) close to the melting temperature of the alloys~\cite{tehranchi2023metastable}, shown in Fig.~\ref{fig:DFTCross}c.
It can be seen that the configurational entropy  yields a domain of chemical potentials, in which the C36 phase becomes stable at $T$ = 773 K.  
This observation is in accordance with the previous studies of Kevorkov et al.~\cite{kevorkov2010400} and Zhong et al.~\cite{zhong2006al2} who found the C36 phase to be stable only at high temperatures. 

The competition of any other bulk phase of interest with C36 can be included in the phase diagram using the same procedure. It can be seen at the top corner of the phase space domain ($\Delta\mu_{\rm Al},\Delta\mu_{\rm Mg}\approx 0$) that the Mg$_3$Al hexagonal phase is also stable at finite temperature only and that the domain of its stability shrinks by decreasing temperature.  Thus we expect that in Al-rich experimental samples, i.e., $\Delta\mu_{\rm Al}\approx 0$ we observe this precipitate, while in Al poor ($\Delta\mu_{\rm Al}<<0$) experimental samples only the C14 phase should be observed.  

However, despite the prediction of the ternary phase diagram, the  C15-CaAl$_2$ Laves phase is not observed in these casts.  
To investigate this discrepancy we again invoke the concept of metastable phase diagrams: The formation of the end member precipitates in the C36 matrix is associated with the formation of interfaces.  Figs.~S3a--b show the structure of the super-cell containing C14/C36 and C15/C36 interfaces, respectively.   The green lines in the ternary phase diagrams along which the formation energy of these interfaces becomes zero are shown  in Fig.~\ref{fig:DFTCross}d--e. It can be seen that the stability region for the formation of the  C15-nucleus, i.e., the grey region, is now shrunken towards the Mg-poor corner of the diagram $(\Delta\mu_{\rm Ca},\Delta\mu_{\rm Al}\approx 0,\Delta\mu_{\rm Mg} \ll 0)$.  In the case of C14, this shrinkage is toward the Al-poor corner.  Taking $\Delta\mu_{\rm Mg} \approx 0$ into account,   
this means that despite the presence of C14, the formation of C15 is suppressed by the barrier configurations.  This manifests the power of metastable phase diagrams to rationalize the experimental observations.

The fact that in sample S1 the precipitates of the hcp-Mg$_3$Al is found, suggests that in this sample the initial chemical potentials were sufficient to overcome the barrier for the formation of this phase.  Thus the coexistence of C36 and hcp-Mg$_3$Al happens at casting.  And given the fact that in both samples the C36 Laves phase is embedded in the Mg matrix, one can find the chemical potential state of this sample, i.e. $\Delta\mu_{\rm Al}=-0.1$eV, $\Delta\mu_{\rm Mg}=0$, and $\Delta\mu_{\rm Ca}=-0.7$eV.  

However, in sample S2 no hcp-precipitate is observed in the C36 matrix.  Thus, the exact values for chemical potentials of Al and Ca remain unclear although since the C14 phase is also not observed one can deduce that $-0.49\le \Delta\mu_{\rm Al}\le 0$ eV, $\Delta\mu_{\rm Mg}=0$, and $\Delta\mu_{\rm Ca}=0.81-1.375\Delta\mu_{\rm Al}$.     
\subsection{Metastable planar defect phase diagrams}\label{sec:MDPD}

To explain the formation of different planar defects in the C36 sample after deformation, we use the concept of defect phase diagrams described in  Ref.~\cite{tehranchi2023metastable}.   The results of the EDS analysis of the deformed sample S1 indicate that the composition of the planar defects is approximately Ca$_{0.45}$Mg$_{1.2}$Al$_{1.35}$.  This elemental composition, which is significantly different from that of C36 phase is close to the average of the elemental composition of the hcp Mg$_3$Al phase and the host C36-CaMg$_{0.625}$Al$_{1.375}$.  Thus it can be envisioned that in the regions close to the precipitates, this Al-rich phase forms due to kinetic reasons during the precipitation process.  
We consider this defect, which extends over the basal plane as the decorated planar defect in our analysis.  On the other hand EDS analysis didn't show any significant segregation on the basal stacking fault which was observed in Sample S2. Therefore, we consider this defect as undecorated.

As mentioned in the previous subsection, we impose $\Delta \mu_{\rm Mg}=0$ according to the fact that the C36 phase is in both samples embedded in a Mg matrix.  This condition is pertinent to the left side of the triangles depicted in Fig.~\ref{fig:DFTCross} and makes the MDPDs one dimensional. 

The formation energy of the undecorated basal defect is calculated using  DFT calculations. The stacking fault energy can be estimated using an extended version of the axial next-nearest neighbor Ising (ANNNI) model~\cite{fisher1980infinitely,selke1988annni,chu1995stacking}, which was previously used for finding the stacking fault energies in C14, and C15. In this model the stacking fault energy is approximated using the lattice parameter of the C36 phase and the energy difference of the corresponding Laves phases.  For the  calculation of the basal stacking fault (BSF) energy, we use the following relation:
\begin{align}
    E_{\rm f}^{\rm BSF}=\frac{10}{\sqrt{3}}\frac{E_{\rm f}^{\rm C14}-E_{\rm f}^{\rm C36}}{a_{\rm C36}^2},
\end{align}
where $E_{\rm f}^{\rm C15}$, $E_{\rm f}^{\rm C36}$, and $a_{\rm C36}$ are the formation energies per atoms of the C15 and C36 phase, as well as the lattice constant of C36, respectively.  The derivation of this formulation is given in section SIV of the Supplementary information.  We note that since the values for $E_{\rm f}^{\rm C15}$, $E_{\rm f}^{\rm C36}$ depend on the spatial distribution of the Al and Mg atoms in the bulk super-cells (see the scatter in Fig.~\ref{fig:DFTCross}a), the stacking fault energy will have a distribution.  The average, maximum and minimum values of basal stacking faults are 4.00, 6.22, and 5.14 meV/$\text{\AA}^2$, respectively.  

\begin{figure}[ht]
         \centering
         \includegraphics[width=\textwidth]{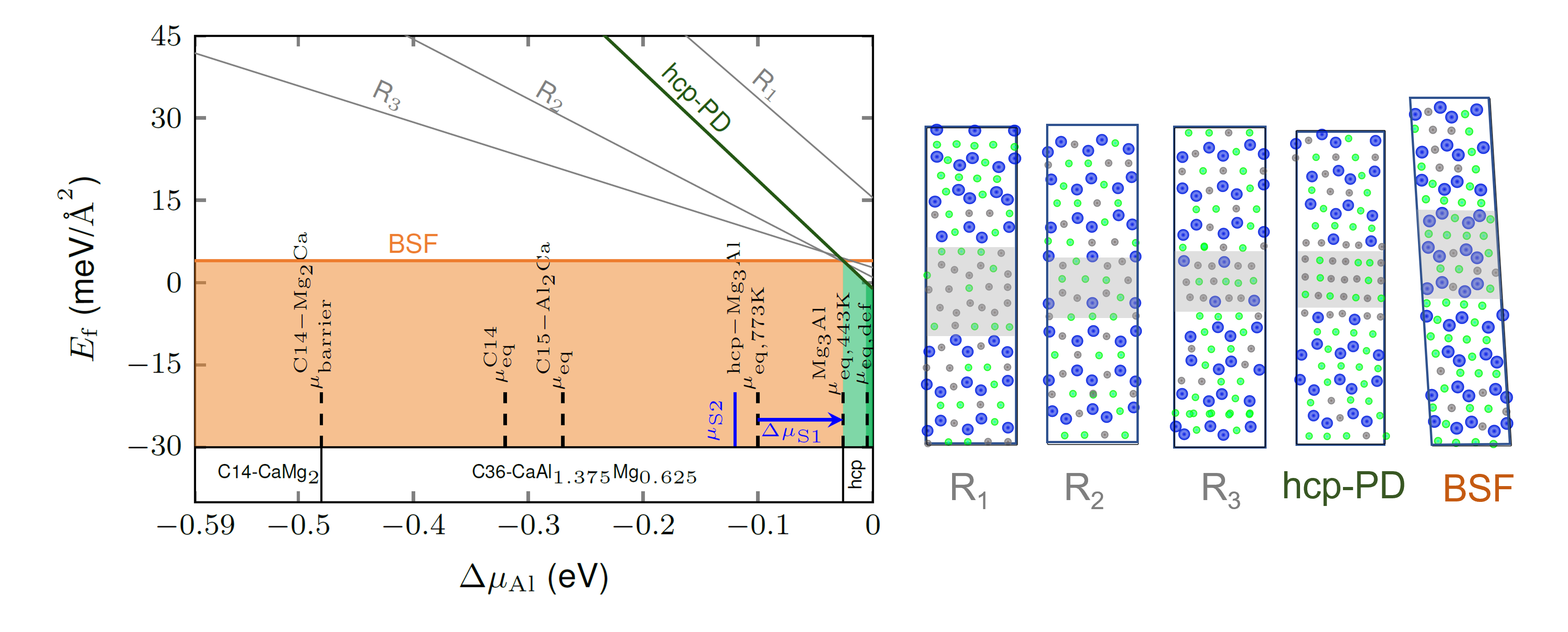}
         \caption{ Formation energies of structures with and without defects are plotted as a function of the chemical potential of Al. The (meta)stable bulk phases pertinent to each chemical potential domain are given in the lower most box.  The color coding is due to the nature of defect formation.  In the orange region the undecorated defect is dominant, whereas in the light green and green regions, the decorated defect is formed by endothermically and exothermically, respectively.  The atomic structures of the basal stacking fault (BSF), the hcp-planar defects,  and other competing defects are given in the right panel.  The blue, green and gray spheres denote Ca, Al, and Mg atoms, respectively.}
         \label{Fig:DPD_hcp}
     \end{figure}

For the decorated defect we start with a super-cell containing the C36 structure with a stacking fault and substitute Ca atoms as the pinning atoms for the triple layer with Mg atoms.  We examine various planar defects in which the concentrations of Mg atoms at the planar defect are 50\%, and 75\%.  The MDPD is given in Fig.~\ref{Fig:DPD_hcp}.
  
Note that sample S1 contains the hcp-precipitates while the C15 phase and C14 phase do not appear in the as cast conditions.  This means that the barrier for formation of this precipitate is overcome during casting and we should not take it into account any more.  The presence of this precipitate fixes the chemical potential of Al at the coexistence of hcp-Mg$_3$Al and C36.  This coexistence only occur at $\Delta\mu_{\rm Al }=-0.1$eV.  At this chemical potential both types of defects are at the endorthermic  region (i.e. their formation energy is positive) but the basal stacking fault is dominant due to its lower formation energy.  However, the deformation is applied at 443K, in this temperature the precipitate is only stable  at $\Delta\mu_{\rm Al }\ge -0.026$eV.  Thus, now the chemical potential of Al can be increased to this value as illustrated by the blue horizontal arrow in Fig.~\ref{Fig:DPD_hcp}.  In this chemical potential both planar defects are still endothermic but now the hcp-planar defect is dominant.  The fact that the positive formation energy of this defect is 3.9 meV/$\AA^2\approx 63$mJ/m$^2$ makes it achievable by application of strain up to 5\% which is the ultimate strain applied to the samples.  

Sample S2 lacks hcp precipitates in addition to bulk phases C15, and C14. As a result, unlike S1, the chemical potential of Al cannot be increased to $-0.026$ eV during deformation. The estimated chemical potential of Al in this alloy at deformation, can be calculated using $\Delta\mu_{\rm Al}=\Delta\mu_{\rm Al}^0+k_{\rm B}T_{\rm deform}\ln c_{\rm}$ where $\Delta\mu_{\rm Al}^0=0.045$eV is the solution enthalpy of Al in Mg~\cite{tehranchi2023metastable}.  This chemical potential, $\Delta{\mu}_{\rm Al}=-0.12$eV, is indicated by the blue vertical line in Fig.~\ref{Fig:DPD_hcp}. At this chemical potential, basal stacking faults remain dominant and, being in the endothermic region, become active during deformation.

It's important to note that decorated and undecorated planar defects have different mechanical behavior under loading. Undecorated planar defects, which are basal stacking faults, can act as carriers of shear slip and promote dislocation formation. On the other hand, decorated planar defects do not have associated stacking faults, so dislocation formation is not promoted. Thus, sample S1 is expected to have a more brittle response compared to S2. This suggests that allowing hcp precipitation to grow can induce a ductile-to-brittle transition in the C36 phase.

\section{Conclusions}\label{Sec:conc}
We have used metastable bulk phase and defect phase diagrams to study the stability of Laves phases and dominant defect phases in Mg-based alloys. DFT calculations were performed to determine the formation energies of bulk phases and of the involved barrier structures. Using this information, we constructed a metastable phase diagram in the chemical potential domain. The results revealed that the formation of the C15 phase is suppressed in Mg-rich samples due to the high interface energy with the C36 phase at the nucleation stage.

We have also calculated the formation energy of planar defects at the basal plane of C36 and have plotted it as a function of the Al chemical potential. In Sample S1, where hcp precipitates formed at 773~K, increasing the Al chemical potential to the co-existence point with C36 at 443~K made the hcp defects dominant but endothermic. In Sample S2, where precipitates were not present, this increase was not possible and basal stacking faults remained dominant. Both samples showed an endothermic formation of defects, which explains why the latter are only observed after deformation.

\section*{Acknowledgement}
The authors acknowledge financial support by the Deutsche Forschungsgemeinschaft (DFG) through the
projects A03, A06, and C05 of the CRC1394 ``Structural and Chemical Atomic Complexity -- From Defect Phase Diagrams to Material Properties'', project ID 409476157.
\bibliography{./bibliography}
\appendix
\end{document}